\documentclass[prl,twocolumn,showpacs,amsmath,amssymb,superscriptaddress,floatfix]{revtex4}

\usepackage[english]{babel}
\usepackage{graphicx}
\usepackage{times}

\begin{document}

\title{Numerical approach  to low-doping regime of the t-J model}

\author{J. \surname{Bon\v ca}}
\affiliation{Faculty of Mathematics and Physics, University of
Ljubljana, Ljubljana, Slovenia} \affiliation{J. Stefan Institute,
Ljubljana, Slovenia}

\author{S. \surname{Maekawa}}
\affiliation{Institute for Materials Research, Tohoku University,
Sendai 980-8577, Japan}
\affiliation{CREST, Japan Science and
Technology Agency (JST), Kawaguchi, Saitama 332-0012, Japan}

\author{T. \surname{Tohyama}}
\affiliation{Yukawa Institute for Theoretical Physics, Kyoto
University, Kyoto 606-8502, Japan}

\date{\today}

\begin{abstract}
We develop an efficient numerical method for the description of a
single-hole motion in the antiferromagnetic background. The method
is free of finite-size effects and allows calculation of physical
properties at an arbitrary wavevector. Methodical increase of the
functional  space leads to results that are  valid in the
thermodynamic limit. We found good agreement  with cumulant
expansion, exact- diagonalization approaches on finite lattices as
well as self-consistent Born approximations. The method allows a
straightforward addition of other inelastic degrees of freedom,
such as lattice effects.  Our results confirm the existence of a
finite quasiparticle weight near the band minimum for a single
hole and the existence of string-like peaks in the single-hole
spectral function.
\end{abstract}


\maketitle

\section{introduction}

A description of hole motion in the antiferromagnetic (AFM)
background as described by the $t-J$ model represents one of the
long-standing, open, theoretical problems in the field of
correlated systems. The accurate solution of this problem may be
crucial for understanding the behavior of high-temperature
superconductors  in the underdoped regime. Apart from the
analytical solution in the Nagaoka regime \cite{nagaoka} in the
limit of zero doping and  small $J/t$, as well as rigorous
theorems in the symmetric point $J=2t$ \cite{sorella1}, there are
no exact solutions of this model in two spacial dimensions. Many
outstanding, early approaches to this problem, such as the
self-consistent Born approximation (SCBA)
\cite{ramsak1,varma,martinez,liu,ramsak98}, self-consistent
perturbational approach (SCPA) \cite{liu}, calculations based on
the string picture \cite{brinkman,trugman,shraiman1},  cumulant
expansion (CE) technique \cite{bonca}, exact diagonalization (ED)
calculations on small clusters \cite{dagotto,poilblanc1,leung},
quantum Monte carlo calculations \cite{sorella},  and recent
state-of-the-art QMC calculations
\cite{mishchenko,brunner,nagaosa} have provided quantitative
description of the quasiparticle band-width, effective mass, and
quasiparticle weight. Most of these methods reproduce dynamical
properties, such as the one-hole spectral function, as well.

Among these  approaches, ED calculations on small clusters provide
exact solutions of the $t$-$J$ Hamiltonian but suffer from
finite-size effects due to small system sizes. Similarly, QMC
calculations are limited to small, even though larger clusters. In
addition, analytic continuation is necessary to obtain spectral
properties, since most of QMC methods compute Green's function,
defined  in imaginary time. The SCBA and SCPA calculations are
likewise limited to finite-size calculations in momentum space.
Furthermore, they seem to overemphasize the string effect. On the
other hand, early calculations based on  the string picture are
similar to the concept of the linear combination of atomic
orbitals, which provide  results for arbitrary momentum transfer.
However, previous results \cite{trugman,shraiman1,inoue} are not
necessarily comparable with SCBA, SCPA and ED results of the
$t$-$J$ model. This is  predominantly due to a limited number of
variational parameters or  to  the small size of the Hilbert space
used in these calculations. In comparison to the variational
approach used in Ref.~\cite{inoue}, where the authors use a
similar method  for construction of the functional basis set, our
method employs an exact-diagonalization approach using Lanczos
technique, which allows solutions of  much larger Hilbert spaces.

The aim of this work is to present an accurate exact
diagonalization method, defined over a limited functional space
(EDLFS). The method is based on the string picture
\cite{brinkman,trugman,shraiman1}, which provides solutions to a
single-hole problem in the AFM background that are free of
finite-size effects.  Furthermore, the method  takes the advantage
of modern computing capabilities that allow solutions of large
matrices. Through the efficient construction of the limited
functional space (LFS), even when using only a few thousand
states, this method provides results that can be directly compared
to the state-of-the-art numerical approaches on small lattices
\cite{leung} that require tens of millions of states.

Despite much work in this area \cite{dagotto_rev} there remain
many open questions concerning the physics of a doped  AFM in the
zero-doping limit. Current interest in this field is in part
focused on the influence of the electron-phonon interaction on
correlated motion of a hole in the AFM  background
\cite{ramsak2,shen,nagaosa,rosch}. Another open question concerns
the proper description of the difference between the hole-  and
electron-doped cuprates \cite{tohyama}. There is also a need for a
method that would resolve the issue of a disappearing
quasiparticle weight that was predicted in a thermodynamic limit
of a doped AFM  due to the phase string effect  \cite{sheng}.

\section{Method}

 We start by writing the $t$-$J$ model as
\begin{eqnarray}
H&=& H_t + H_{\|} + H_{\perp} \\
H_t&=& -t \sum_{\langle{\bf  i,j} \rangle,s}\tilde c^\dagger_{{\bf
i},s}
\tilde c_{{\bf j},s} + {\mathrm{h.c.}}\nonumber\\
 H_{\|}&=& J\sum_{\langle \bf i,j\rangle }S^z_{\bf i} S^z_{\bf
   j}\nonumber \\
 H_{\perp}&=& J/2\sum_{\langle \bf i,j\rangle}S^+_{\bf i} S^-_{\bf j} +
 S^-_{\bf i} S^+_{\bf j}, \nonumber
 \label{ham}
\end{eqnarray}
where $\tilde c_{{\bf i},s}=c_{{\bf i},s}(1-n_{{\bf i},-s})$ is a
fermion operator, projected onto a space of no double occupancy.
The  sums run over the pairs of nearest neighbors as well as over
the two spin orientations. Following in part  works by  Trugman
\cite{trugman}, Inoue and Maekawa \cite{inoue}, and  El Shawish
and J.B. \cite{samir}, we construct the LFS starting from a
N\'{e}el state with one hole, $\vert \phi_{\bf k}^{(0)}\rangle  =
c_{\bf k}|{\mathrm{Neel}}\rangle$ and proceed with  generation of
new states, by application of  the kinetic part of the Hamiltonian
$H_t$, {\it i.e.}
\begin{equation}
\{ \vert \phi_{{\bf k}l}^{(N_h)}\rangle \}=H_t^{N_h}\vert
\phi_{\bf k}^{(0)}\rangle.\label{generate}
\end{equation}
This procedure generates  strings with maximum lengths given by
$N_h$. While constructing  the LFS, translation symmetry,
generated by two minimal translations ${\bf r_{1,2}}=(1,\pm 1)$,
is taken into account. Due to exponentially rapid growth of the
LFS with increasing $N_h$ we  introduce an additional parameter
$N_b\leq N_h$ that restricts generation of long strings by
imposing a  condition under which   all coordinates of spin-flips
should satisfy $\vert \mu_h-\mu_f\vert \leq N_b; \mu=\{x,y\}$
where $h$ and $f$ refer to electron and spin-flip indexes,
respectively. Application of this condition improves the quality
of the LFS by increasing the number of states containing
spin-flips in the vicinity of the hole while keeping the total
amount of states within  computationally accessible  limits. The
full Hamiltonian in Eq.~\ref{ham} is then diagonalized within this
LFS using the standard Lanczos procedure. We also note that our
method, even though defined on the infinite lattice, is
variational. Increasing the number of LFS systematically lowers
energies of the zero- and single- hole states, {\it i.e.} $E^{0h}$
and $E^{1h}$.

While generation of single-hole states through application of only
the kinetic part of the Hamiltonian seems a rather natural choice
for the construction of the single-hole wavefunction, there
remains a question of how to construct the LFS for the undoped
case, {\it i.e.},  the Heisenberg model. The solution of the
latter  seems necessary in order  to compute spectral properties
of the one-hole system as well as its energy, relative to the
undoped case. We next assume that the spacial extent of the
disturbance of the spin background around the doped hole (in the
literature also referred as a {\it magnetic polaron}) is finite.
In this case it is not necessary to obtain the exact solution of
the undoped system on the infinite 2D lattice for the correct
description of the single hole properties. It is sufficient to
find a solution of the Heisenberg model in the vicinity of the
doped hole. We therefore construct the 0-hole LFS using the 1-hole
LFS  by simply filling the empty space with a spin.

\begin{figure}[htb]
\includegraphics[width=8cm,clip]{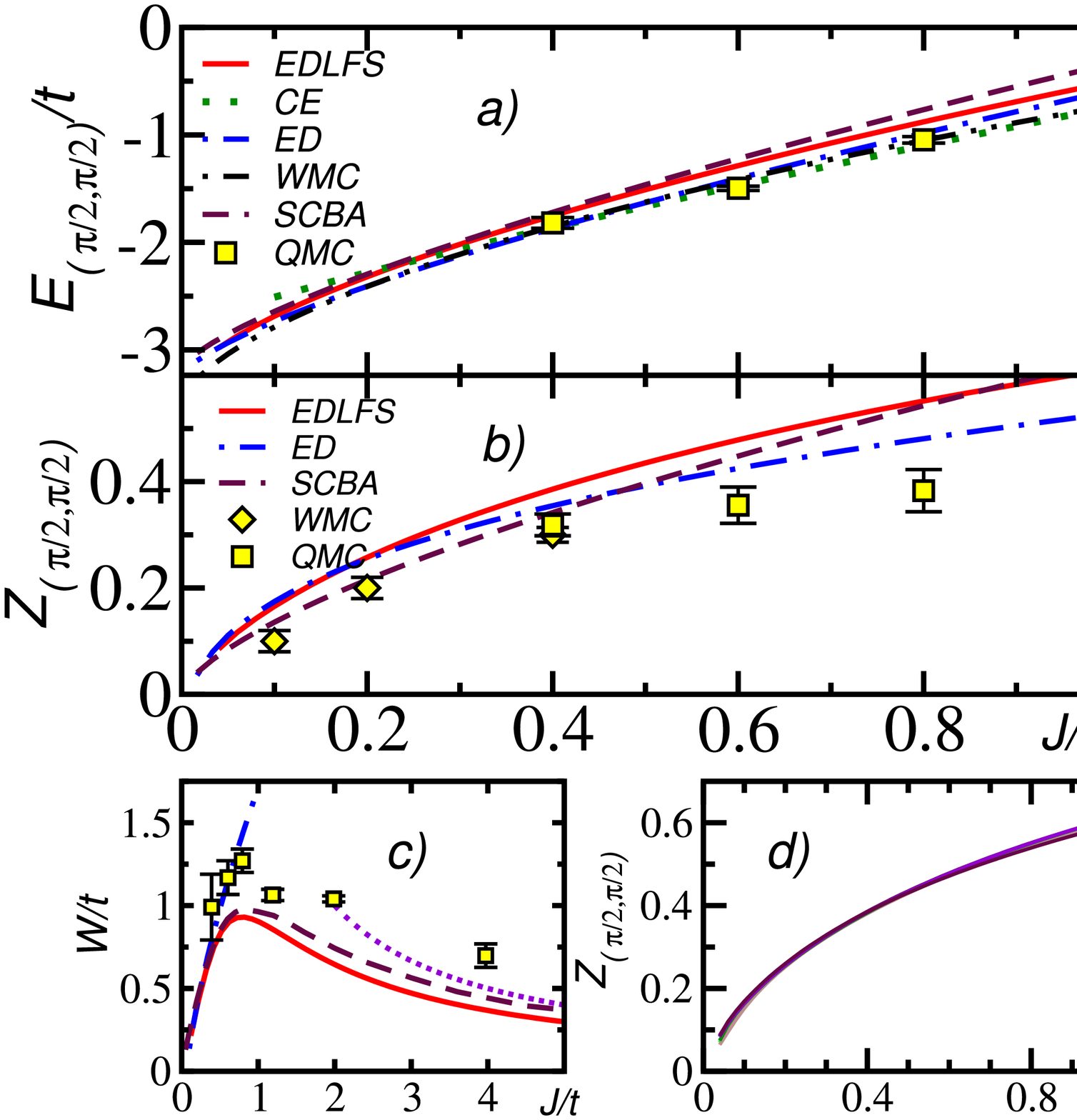}

\caption{(Color online) a) Single-hole energy $E_{{\bf
k}=(\pi/2,\pi/2)}$ vs. $J/t$. The full line represents EDLFS results
of the present work that were obtained using $N_{st}=37402972$
states. Dotted line represent CE results from Ref. \cite{bonca}. The
dot-dashed line, given by $E_{{\bf k}=(\pi/2,\pi/2)}=J/t-3.24+ 2.65
(J/t)^{0.72}$ is an interpolation based on ED results obtained on 32
sites \cite{leung}. The double-dot-dashed line represents
interpolation $E_{\bf k=(\pi/2,\pi/2)}=-3.36+ 3.50 (J/t)^{2/3}$ based
on WMC method from Ref.~\cite{mishchenko}. The dashed line, $E_{\bf
k=(\pi/2,\pi/2)}=-3.17+ 2.83 (J/t)^{0.73}$ is a result of SCBA
calculation, \cite{martinez,szcz}.  Squares represent QMC calculations
from Ref.~\cite{brunner}. Note that $E_{\bf k}/t$ results of ED, WMC
and QMC calculations were shifted by $-J/t$ due to a different
deffinition of the $t$-$J$ model.
%
%
b) Quasiparticle weight $Z_{{\bf k}=(\pi/2,\pi/2)}$ vs. $J/t$. The
full line represents EDLFS calculation  using $N_{st}=37402972$
states. The  dot-dashed line represents ED results as summarized
in the extrapolation given by $Z_{{\bf k}=(\pi/2,\pi/2)}=-0.136 +
0.664 (J/t)^{0.333}$ \cite{leung}, while the  dashed line
represents SCBA calculation $Z_{{\bf k}=(\pi/2,\pi/2)}=0.63
(J/t)^{0.667}$ from Ref.~\cite{martinez} (similar extrapolation
was also obtained using SCPA in Ref.~\cite{liu}). Diamonds
represent WMC calculation from Ref.~\cite{mishchenko}, and squares
are QMC results from Ref.~\cite{brunner}. Note that WMC and QMC
data were multiplied by a factor of 2 due to a different
definition of $Z_{{\bf k}}$ used in
Refs.~\cite{brunner,mishchenko}.
c) The bandwidth $W/t$ vs. $J/t$ calculated with EDLFS (full
line), ED \cite{leung} (dot-dashed line), SCBA
\cite{martinez}(dashed lines), QMC calculations from
Ref.~\cite{brunner} (squares),  and $W/t=2t^2/J$ (dotted line),
proposed in Ref.~ \cite{martinez}.
d)$Z_{\bf k}$ vs. $J/t$ (ten nearly overlapping curves) obtained
by using 10 different sizes of LFS as explained in the text.
}\label{fig1}
\end{figure}

\begin{figure}[htb]
\includegraphics[width=8cm,clip]{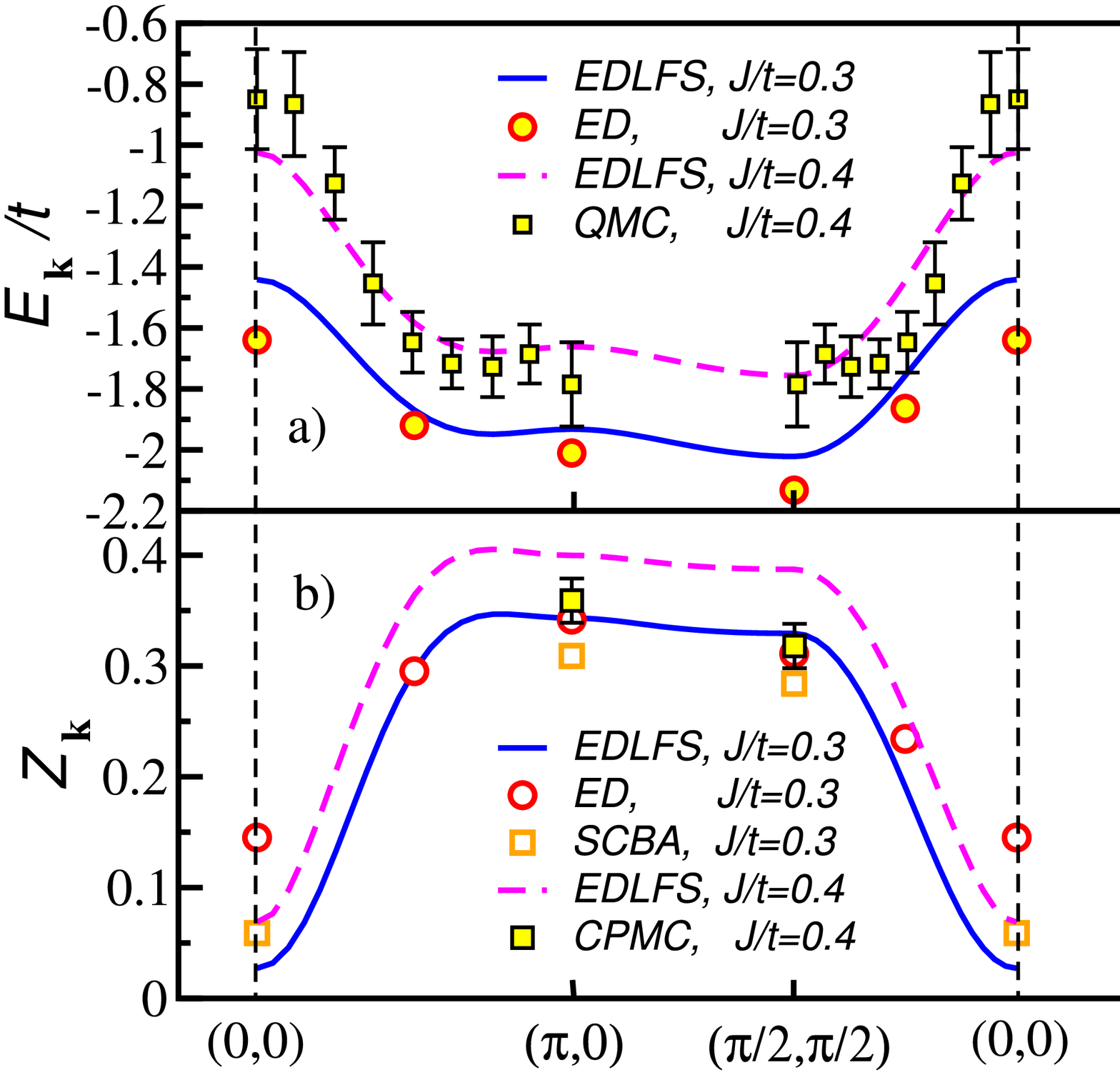}
\caption{(Color online)  a) Single-hole energy $E_{\bf k}$ and b)
quasiparticle weight $Z_{\bf k}$. Both quantities were calculated
at fixed $J/t=0.3$ and 0.4 as noted in the figure. EDLFS results
of  the present work were obtained with $N_{st}=37402972$ states
(full lines), ED results  (circles), calculated on a 32 site
system, are from Ref.~\cite{leung},  QMC results (squares) are
from Ref.~\cite{brunner}, while  SCBA results (diamonds)  in b)
are from Ref.~\cite{martinez}.}
\label{fig2}
\end{figure}

\section{Static properties}

We now turn to the numerical results. The one-hole energy,
measured from the energy of the undoped system, $E_{\bf k} =
E^{1h}_{\bf k}-E^{0h}$, is shown at the one-hole band-minimum
${\bf k}=(\pi/2,\pi/2)$ in Fig.~\ref{fig1}a, along with CE
\cite{bonca}, ED \cite{leung}, worm quantum Monte Carlo (WMC)
\cite{mishchenko,nagaosa}, Quantum Monte Carlo (QMC) calculations
\cite{brunner}, and SCBA \cite{martinez,szcz} results. While our
method is defined on the infinite system, the absolute values of
$E^{1h}_{\bf k}$ and $E^{0h}$ are ill-defined since they grow with
the increasing number of basis states as the number of spin-flips,
generated by the hole motion, increases. In contrast, $E_{\bf k}$
remains finite and well defined. Although our method can not be
directly compared to the cumulant expansion technique
\cite{bonca}, we use some of the aspects of this technique. Since
we use in our method only the hopping part of the Hamiltonian,
Eq.~\ref{ham}, to generate new states, all spin-flips are by
construction limited to the vicinity of the hole. This by no means
restricts the LFS only to connected strings. A propagating hole
can also generate disconnected strings. In our approach, the
precision of the description of the quantum spin fluctuations,
caused by the presence of the hole, increases with decreasing
distance from the hole. We can therefore expect to achieve a
thermodynamic limit as soon as the extent of the spin-flips in the
LFS exceeds the size of the magnetic polaron. In addition, we
should stress that the zero-hole energy, $E^{0h}$, {\it per se}
has no physical meaning. It simply represents the solution of the
Heisenberg model, defined on the zero-hole LFS, that is identical
to the one-hole LFS  with the exception of the additional spin
located on the hole position. The high efficiency of our approach
is reflected in good agreement of our results with CE method,
\cite{bonca}, and SCBA approach \cite{martinez,szcz}. For
comparison we also present results obtained with ED calculation on
a 32 site system \cite{leung}, WMC calculations \cite{mishchenko},
as well as with QMC calculations \cite{brunner} performed  on much
larger lattices (24x24) (see Fig.~\ref{fig1}a). In general, EDLFS,
CE as well as SCBA methods give consistently lower values of  the
single-hole (polaron) energy in comparison to ED and QMC methods.
Here we stress, that the single-hole energy is extremely sensitive
to the appropriate choice of the LFS for the 1- as well as of the
0- hole space. Our results can be almost perfectly fitted with a
form $E_{(\pi/2,\pi,2)}/t =\omega_0= a_0 + b_0 (J/t)^{\gamma_0}$
where parameters $a,b$ and ${\bf k}=0$ are listed in the first row
of Table~\ref{tab_fit}.


We present the bandwidth $W$ in Fig.~\ref{fig1}c along with SCBA
\cite{martinez} and QMC results \cite{brunner},  as well as with
analytical prediction $W=2 t^2/J$ \cite{martinez}, valid in the
large $J/t$ limit. We find good agreement with ED results in the
physically most relevant regime $J/t\sim 0.4$.  We note that in
Ref.~\cite{leung} $W$ is defined as $W= E_{(\pi,\pi)}-
E_{(\pi/2,\pi/2)}$. In our approach due to broken translation
symmetry the point ${\bf k}=(\pi,\pi)$ is folded onto the ${\bf
k}=0$ point. We thus believe that our definition of $W$ is
comparable to the one in Ref.~\cite{leung}. QMC results from
Ref.~\cite{brunner} in contrast to EDLFS, ED and SCBA results
predict slightly larger values of $W/t$.  Note however larger
error bars in QMC results around $J/t\lesssim 0.6$.

\begin{figure}[htb]
\includegraphics[width=8cm,clip]{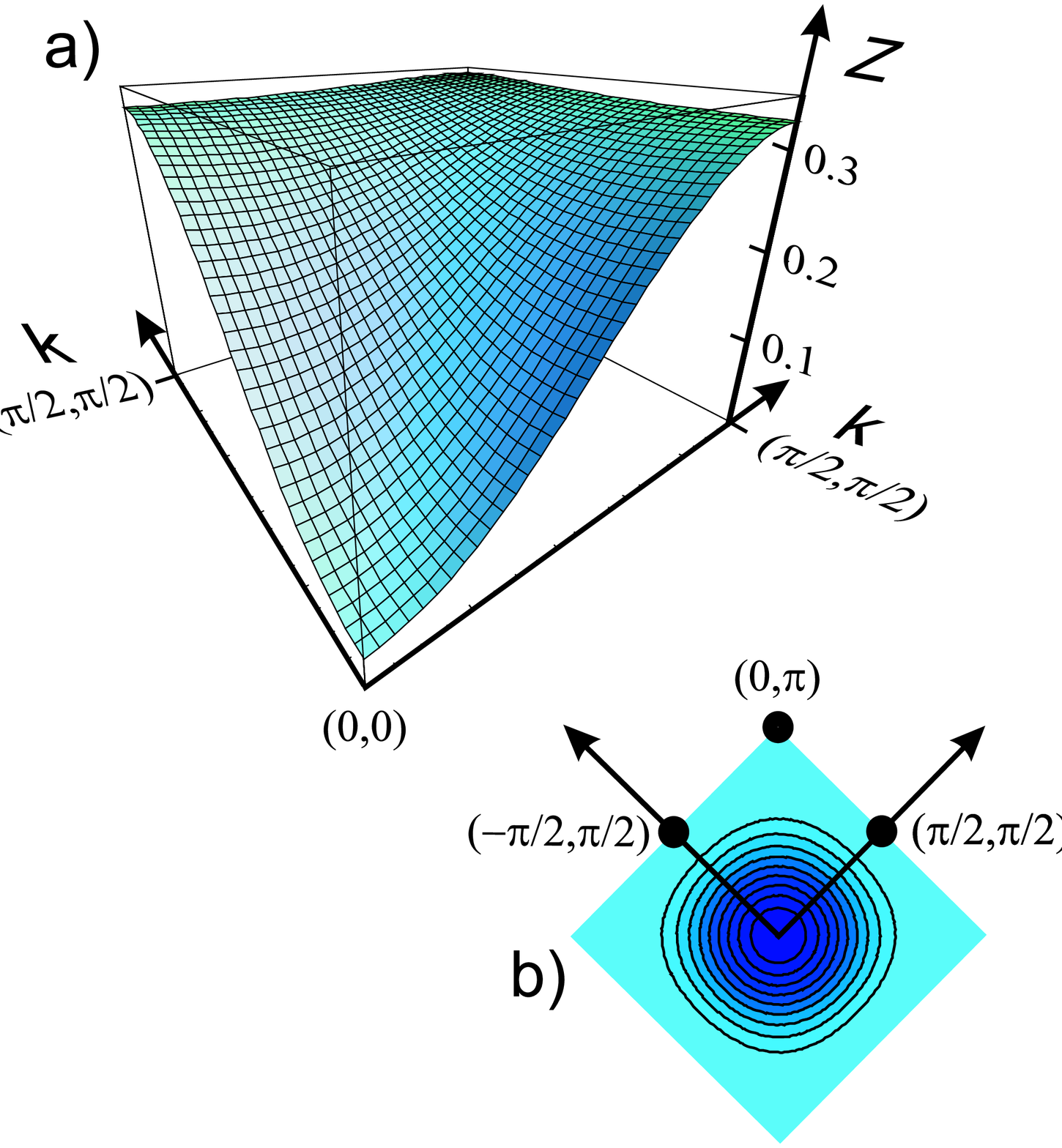}
\caption{(Color online)  a) Surface plot of $Z_{\bf k}$ at
$J/t=0.3$, computed on a mesh of 400 $\bf k-$ points. Variational
space with $N_{st} = 5213618 $ was used. Only 1/4 of the AFM BZ is
shown.  b) Contour plot of  $Z_{\bf k}$ with equidistant contours
while wavevectors are  spanning the whole AFM BZ. The rest is the
same as in a).}
\label{fig3}
\end{figure}

Our calculation of $E_{\bf k}$ presented in  Fig.~\ref{fig2}a
reflects another important  advantage of the present method over
ED calculations on limited system sizes. Note, however, that our
calculations are limited to the reduced AFM Brillouin zone (BZ)
because of  broken translational symmetry. Defining  the LFS on an
infinite lattice  allows  calculation of physical properties at an
arbitrary wavevector, limited to AFM BZ. In Fig.~\ref{fig2}a, we
present the dispersion relation $E_{\bf k}$, calculated at
$J/t=0.3$ and 0.4. Taking into account the fact that we are
computing $E_{\bf k}$ in absolute units (we used no additional
energy shift), we find good agreement with ED results
\cite{leung}, in particular when comparing the ${\bf
k}-$dependence of the single-hole energy and the bandwidth. We
find qualitative agreement also with QMC results from
Ref.~\cite{brunner}, calculated at $J/t=0.4$. QMC method predicts
larger bandwidth as also seen in Fig.~\ref{fig1}c. To further
quantify the efficiency of our method, we present our results for
$E_{\bf k}$ at selected $\bf k-$ points in Table~\ref{table}.
These results were obtained using different numbers of states
spanning the LFS at the physically relevant value $J/t=0.3$. It is
encouraging that reasonable results for the one-hole energy can be
obtained from a LFS as small as $N_{st}=1121$.

So far, we have shown that our method is  successful in obtaining
the ground-state energy of the magnetic polaron, however, the
current scientific interest and existing open problems primarily
concern dynamic properties of a doped hole. Before moving to
dynamic response, we next present  our results of a closely
related quantity, {\it i.e.} the quasiparticle weight, $Z_{\bf
k}$, vs. $J/t$, see Fig.~\ref{fig1}b. We define $Z_{\bf k}$ by
\begin{equation} Z_{\bf k}={ \vert \langle \Psi_{{\bf
k},0}^{1h}\vert c_{\bf k}\vert \Psi^{0h}\rangle\vert^2 \over
\langle \Psi^{0h}\vert c_{\bf k}^\dagger c_{\bf k}\vert
\Psi^{0h}\rangle }, \label{zk}
\end{equation}
where $\Psi_{{\bf k},0}^{1h}$ $(\Psi^{0h})$ represent the ground
state of the system with either one or zero holes. Note that the
ground state $\Psi^{0h}$ has ${\bf k}=0$.  The agreement with the
ED calculation is surprisingly good for $J/t\lesssim 0.3$. The WMC
calculation from Ref.~\cite{mishchenko} and  QMC calculations from
Ref.~\cite{brunner} yield slightly smaller values for $Z_{\bf k}$.
It is noteworthy mentioning that the two different QMC methods
yield consistent values of $Z_{\bf k}$ (note nearly perfect
overlap between the two methods at $J/t=0.4$). We have also tested
the $J/t\to \infty$ limit where exact result based on the
spin-wave approximation  from Ref.~\cite{malshukov} yields
$Z=0.822$.  Our method gives $Z=0.926$, which can be further
compared with SCBA result calculation that gives $Z=1$.

We next briefly discuss possible sources of errors affecting
results obtained by different approaches. In case of SCBA
calculations, the error is due to the approximate nature of the
calculation since only non-crossing diagrams are taken into
account. ED calculations are limited to small lattice-sizes that
may lead biased  results due to finite-size effects. QMC
simulation from Ref.~\cite{brunner}, based on the loop-cluster
Monte carlo method for the AFM state and the hole propagation
within the fixed spin background, yields increasing larger error
bars as one approaches the physically relevant regime $J/t\lesssim
0.6$, while WMC method \cite{mishchenko} suffers from he
minus-sign problem. EDLFS naturally depends on the choice of the
LFS. Increasing the number of LFS should yield results that are
free of finite-size effects and valid in the thermodynamic limit.
Nevertheless, a systematic error may occur due to a particular
algorithm used to create different LFS, Eq.~\ref{generate}. To
demonstrate the stability of our results against the choice of
different LFS, as well as a rapid convergence of our method for
$0.02\lesssim J/t<1$ with increasing $N_{st}$, we present in
Fig.~(\ref{fig1}d) nine nearly overlapping curves depicting
$Z_{\bf k}$. The curves were calculated using different LFS's
with: $N_{st}=7610, 9786, 43884,  80108, 218950$, $642406$,
$912478$, $3109626$ and $5213618$ as obtained using LFS generator,
Eq.~\ref{generate} with various values of $N_h$ and $N_b$. The
close agreement of values for $Z_{\bf k}$ given in
Table~\ref{table} represents additional qualitative demonstration
of convergence in our calculation. Note that results are only
weakly dependent upon the choice of  parameters $N_h$ and $N_b$
that define the generating algorithm for LFS.

\begin{table}[t]\begin{center}
\makebox{\vbox{\hrule depth 0.8pt \hbox{\vspace{-1pt}}
\hbox{\begin{tabular}{|c|ccc|}
 & $a_n$ & $b_n$ & $\gamma_n$ \\
 \hline

$\omega_0$ & -3.37  & 2.86  & 0.62  \\[0.2cm]
$\omega_1$ & -3.39  & 4.50  & 0.76  \\[0.2cm]
$\omega_2$ & -3.12  & 5.56  & 0.72  \\[0.2cm]

\end{tabular}}
\hrule depth .8pt}}
\caption{Fitting parameters  of the lowest peak positions in
$A_{(\pi/2,\pi/2)}(\omega)$. Fits have the from $\omega_n = a_n +
b_n (J/t)^{\gamma_n}$. } \label{tab_fit}
\end{center}\end{table}

In Fig~\ref{fig2}b we  present $Z_{\bf k}$ along the special
symmetry lines in the reduced AFM BZ. The agreement with the ED
result is good for  large values of $k=\vert {\bf k}\vert$ while
the agreement with SCBA calculation is poorer. The discrepancy
between our method and SCBA  is  similar over the whole AFM BZ
since the SCBA does not suffer from finite-size effects. Most
importantly, we find the value of $Z_{\bf k}$ to be very small
around the ${\bf k}=0$ point (see also Table~\ref{table}),
followed by a sharp increase with increasing $k$. These
observations are consistent with the SCBA result.

A surface plot of $Z_{\bf k}$ in Fig.~\ref{fig3}a that consist of
400 $k$-points, calculated on a system with $N_{st}= 5213618$
states, shows the power of our method.  As expected from results,
plotted in Fig.~\ref{fig2}, $Z_{\bf k}$ shows a pronounced minimum
located at ${\bf k}=0$ followed by a rapid increase with
increasing distance from the ${\bf k}=0$ point. In
Fig.~\ref{fig3}b we show contour plot of $Z_{\bf k}$ over the
whole AFM BZ where contour lines, representing values of $Z_{\bf
k}$, are uniformly spaced in the interval $[0,0.3]$.

\begin{figure}[htb]
\includegraphics[width=8cm,clip]{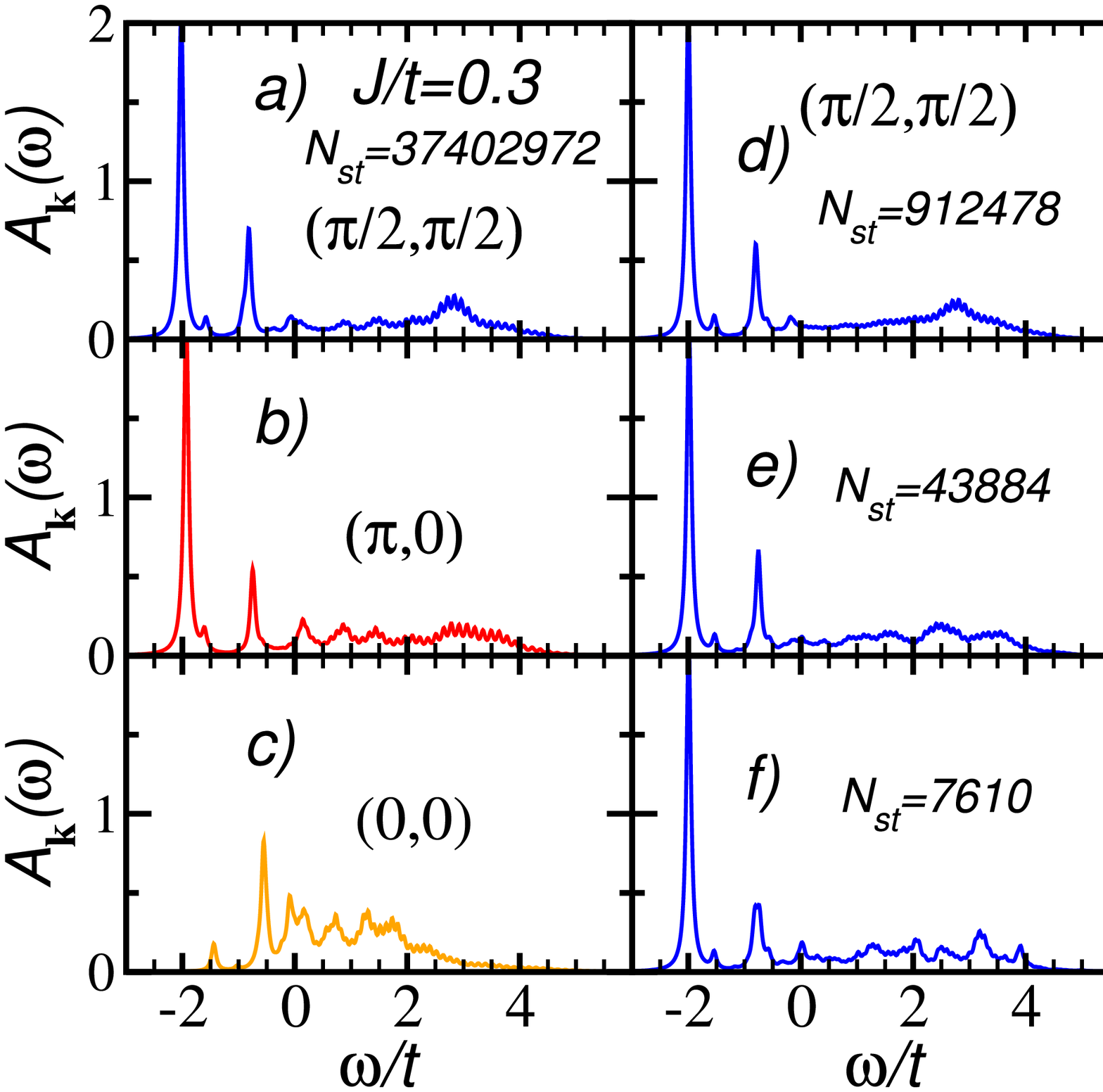}
\caption{(Color online)  a-c) Spectral functions $A_{\bf
k}(\omega)$ for three typical  values of ${\bf k}$, d-f) $A_{\bf
k}(\omega)$ at ${\bf k}=(\pi/2,\pi/2)$ calculated using different
sizes of LFS's as indicated in the insets.  In all cases, we have
used $J/t=0.3$ and an artificial damping of $\epsilon=0.05$. }
\label{fig4}
\end{figure}
\begin{table*}[htb]\begin{center}
\makebox{\vbox{\hrule depth 0.8pt \hbox{\vspace{-1pt}}
\hbox{\begin{tabular}{|c|c|c||ccc||ccc|}
\raisebox{-3mm}{$N_h$} &\raisebox{-3mm}{$N_b$} &\raisebox{-3mm}{$N_{st}$} &
\multicolumn{3}{c||}{\rule[3mm]{0mm}{2mm}$E_{\bf k}$}
& \multicolumn{3}{c|}{$Z_{\bf k}$}\\
&&& \rule[-2.5mm]{0mm}{2mm} $(\pi/2,\pi/2)$ & $(\pi,0)$ & $(0,0)$ &
$(\pi/2,\pi/2)$ & $(\pi,0)$ & $(0,0)$
  \\ \hline\hline
  \rule[4mm]{0mm}{2mm}
6 & 4&  1121 &-2.01925 &  -1.95213 &  -1.44065 &   0.29253 &   0.32780 &   0.00002 \\[0.2cm]
8 & 4& 7610 &-1.99475 &  -1.92799 &  -1.47960 &   0.32617 &   0.33895 &   0.03093 \\[0.2cm]
8 & 8& 9786 &-1.99951 &  -1.92888 &  -1.47982 &   0.32349 &   0.33803 &   0.03077 \\[0.2cm]
10 & 4& 43884 &-1.98751 &  -1.92209 &  -1.45354 &   0.32529 &   0.35097 &   0.03359 \\[0.2cm]
10 & 8& 80108 &-2.00182 &  -1.92305 &  -1.45542 &   0.32486 &   0.34104 &   0.03098 \\[0.2cm]
12 & 4& 218950 &-2.00272 &  -1.93757 &  -1.46192 &   0.32790 &   0.34895 &   0.03535 \\[0.2cm]
12 & 8& 642406 &-2.01059 &  -1.92709 &  -1.43991 &   0.32633 &   0.34345 &   0.03009 \\[0.2cm]
14 & 4& 912478 &-2.00024 &  -1.93322 &  -1.47915 &   0.32902 &   0.34942 &   0.03907 \\[0.2cm]
14 & 8& 4992876 &-2.01830 &  -1.93175 &  -1.44255 &   0.32805 &   0.34314 &   0.02809 \\[0.2cm]
14 & 12& 5225818 &-2.01831 &  -1.93175 &  -1.44255 &   0.32804 &   0.34314 &   0.02809 \\[0.2cm]
16 & 8 & 37402972 & -2.02175 &  -1.93205 &  -1.44112 &   0.32939 &   0.34324 &   0.02713 \\[0.2cm]

\end{tabular}}
\hrule depth .8pt}}
\caption{$E_{\bf k}$ and $Z_{\bf k}$, calculated for $J/t=0.3$ and
different sizes of the LFS, generated using different values of
$N_h$ and $N_b$.}
 \label{table}
\end{center}\end{table*}

\section{Spectral functions}

In Figs.~\ref{fig4}a-c we plot the hole spectral function $A_{\bf
k}(\omega)$, calculated at $J/t=0.3$ using  three typical values
of $\bf k$. We define $A_{\bf k}(\omega)$ as
\begin{equation}
A_{\bf k}(\omega)=\sum_n\vert \langle \Psi_{{\bf k},n}^{1h}\vert
c_{\bf k}\vert \Psi^{0h}\rangle\vert^2 \delta\left (\omega - \left
(E_{{\bf k},n}^{1h}-E^{0h}\right )\right), \label{akom}
\end{equation}
where $ \vert \Psi_{{\bf k},n}^{1h}\rangle $ and $E_{{\bf
k},n}^{1h}$ represent excited states and energies of the 1-hole
system. In many respects,  our results agree with the ED
calculations of Leung and Gooding, Ref.~\cite{leung}. The
quasiparticle peak is well defined for wavevectors lying on the
edge of the AFM BZ. In particular,  for ${{\bf
k}_1}=(\pi/2,\pi/2)$ the peak is  located at $\omega =\omega_0=
E_{\bf k_1}$, see also Table~\ref{table}. The quasiparticle peak
is well defined also for ${{\bf k}_2}=(\pi,0)$. In contrast to ED
results, we see a tiny peak at ${{\bf k}_1}$, located at
$\omega_1\sim -1.58$ that scales with $J/t$ as $\omega_1=a_1 + b_1
(J/t)^{\gamma_1}$, Table~\ref{tab_fit}. This fit is valid  in the
regime $0.3\lesssim J/t\lesssim 1.0$. This peak can also be
distinguished at ${{\bf k}_2}$.   At yet higher frequencies there
is another well defined peak. It is located at $\omega_2\sim
-0.81$ at ${{\bf k}_1}$ with the following scaling  $\omega_2=a_2
+ b_2 (J/t)^{\gamma_2}$, Table~\ref{tab_fit}. This fit is valid
in the regime $0.1\lesssim J/t\lesssim 1.0$. ED results from
Ref.~\cite{leung} also display a well defined but less sharp
structure at these frequencies.  Moving towards ${{\bf k}_2}$,
this peak looses some weight, however, it remains well defined.
Spectrum at larger $\omega$ is broad and mostly featureless. We
note  different scaling with $J/t$ between the quasiparticle peak,
located at $\omega_0$ and string-like peaks, positioned at
$\omega_1$ and $\omega_2$, see Table~\ref{tab_fit}.

At ${\bf k}_3=0$, $A_{{\bf k}_3}(\omega)$ displays a much smaller
quasiparticle peak at $\omega \sim E_{{\bf k}_3}$ than found in ED
calculations. The broad, mostly incoherent part moves to lower
frequencies and slightly shrinks. Nevertheless, the incoherent
structure is broader than in ED calculations.



In Figs~\ref{fig4}a,d-e we plot $A_{{\bf k}_1}(\omega)$ for 4
different sizes of the LFS, ranging from $N_{st}=37402972$ down to
7610. A filling up of the incoherent part of the spectrum in the
large $\omega$ interval $0\lesssim \omega/t \lesssim 4$ is the
predominant effect of increasing $N_{st}$. All special features in
the range $\omega\lesssim 0$ seem to be well captured  within the
smallest size LFS. In addition, spectral functions for the largest
two LFS's (see Figs~\ref{fig4}a and d) nearly overlap in the whole
$\omega / t$ regime.
\begin{figure}[htb]
\includegraphics[width=10cm,clip]{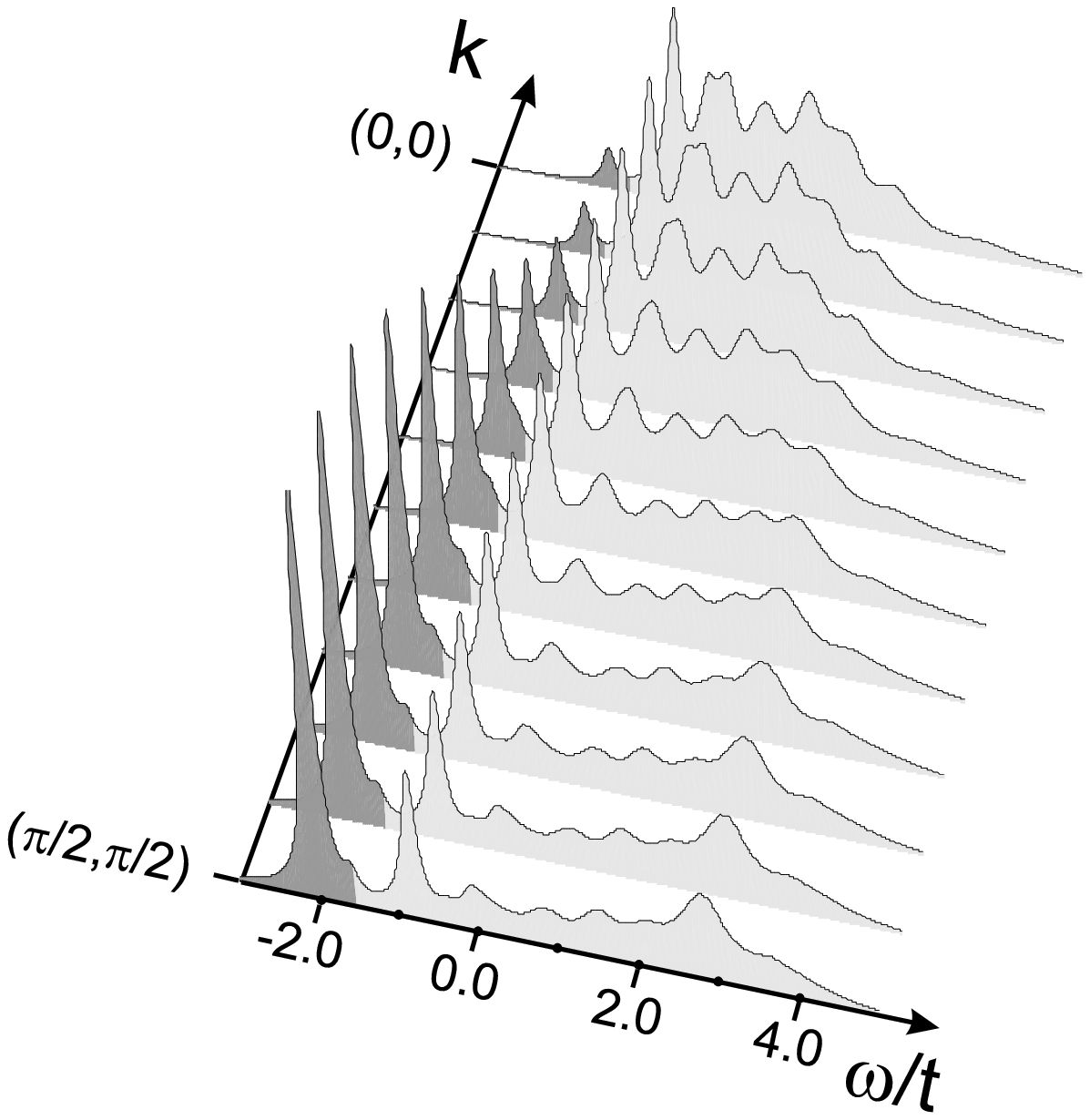}
\caption{ Spectral functions $A_{\bf k}(\omega)$ for wavevectors
${\bf k}=(\pi/2,\pi/2)$ through $(0,0)$.  In all cases we have
used $J/t=0.3$, $N_{st}=5213618$ and artificial damping
$\epsilon=0.1$. Dark-shaded areas are proportional to the
quasiparticle weight $Z_{\bf k}$.}
\label{fig5}
\end{figure}
\begin{figure}[htb]
\includegraphics[width=8cm,clip]{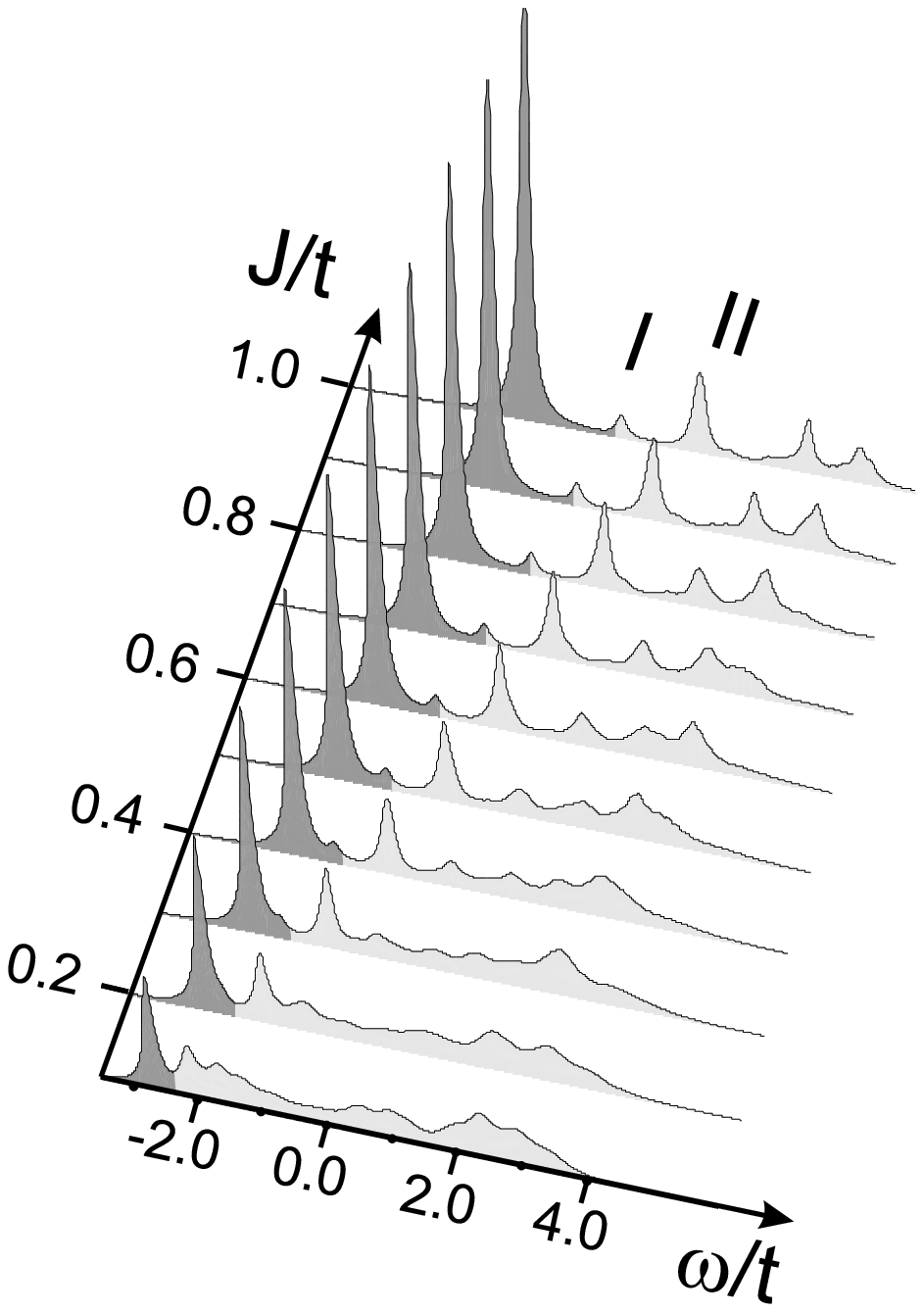}
\caption{ Spectral functions $A_{{\bf k}=(\pi/2,\pi/2)}(\omega)$
for various values of $J/t$ are shown. We have used
$N_{st}=5213618$ and artificial damping $\epsilon=0.1$.
Dark-shaded areas are proportional to the quasiparticle weight
$Z_{{\bf k}=(\pi/2,\pi/2)}$.}
\label{fig6}
\end{figure}

In Fig~\ref{fig5}, we present the evolution of $A_{\bf k}(\omega)$
for ${\bf k}$ moving from $(\pi/2,\pi/2)$ towards $(0,0)$. All
curves were computed using  $N_{st}=5213618$. Dark-shaded  areas
are graphic representations of $Z_{\bf k}$.  The evolution of
spectral functions with increasing values of $J/t$, calculated at
${\bf k}=(\pi/2,\pi/2)$, is presented  in Fig.~\ref{fig6}. The two
lowest string-like peaks are denoted with Roman numerals. Scaling
as discussed in the beginning of this section  of the
quasiparticle peak, and two lowest string-like peak positions,
$\omega_0,\omega_1$, and $\omega_2$,  with $J/t$ can be
qualitatively followed. With increasing $J/t$, the broad continuum
at high-$\omega/t$ transforms into well defined peaks.

\section{Conclusions}

In conclusion, we have developed an efficient numerical approach
for calculating physical properties of a doped AFM insulator in
the zero-doping limit. The presented method is highly efficient,
free of finite-size effects, and it allows for computation of
physical properties at an arbitrary wavevector.
EDLFS obviously has a few  shortcomings: a) the method is limited
to calculations in the zero-doping limit, b) due to the broken
symmetry of the starting wavefunction, calculations are limited to
the reduced AFM BZ, and c) results depend on the number of states
$N_{st}$ spanning the LFS. However,
 for most static as well as dynamic quantities convergence
to the thermodynamic limit with increasing $N_{st}$ can be
achieved.

Using EDLFS, we have computed the quasiparticle energy,
quasiparticle weight, and spectral functions and compared values
to known and established analytical as well as numerical results.
We found the best agreement with CE and SCBA calculations for the
single-hole energy while  for the quasiparticle weight at
$J/t\lesssim 0.3$ best agreement was found with ED calculations
obtained from the largest system of 32 sites. Our method with an
already small number of LFS produces results for static
quantities, such as the energy dispersion, the bandwidth and the
quasiparticle weight, that are directly comparable to the
state-of-the-art ED calculations on small lattices.  Our
simulations show that the quasiparticle weight around the
band-minimum ${\bf k}=(\pi/2,\pi/2)$ remains finite in the
thermodynamic limit. The quasiparticle peak is separated by a
pseudo-gap from  well defined string-like peaks. Comparing our
results to ED calculations, we find a much smaller quasiparticle
weight at the ${\bf k}=0$ point.

Our method can be easily extended to compute other static as well
as dynamic quantities,  {\it e.g.}, various correlation functions
in the vicinity of the doped hole and optical conductivity.
Furthermore, it allows for the inclusion of additional
higher-order terms in the Hamiltonian, such as  the next-nearest
neighbor hopping term that  allows comparison  of hole vs.
electron doped AFM systems. The method can be easily extended to
computation of bound two-hole  properties by adding another hole
to the LFS. EDLFS can also be adopted to computing single-hole
properties of the $t-J$ model on the triangular lattice. In this
case a different, 120$^o$ ordered Neel state, should be used for
the starting wavefunction, Ref.~\cite{trumper}  Finally, by
adopting the method of Ref.~\cite{bonca1,bonca2}, the present
approach offers a natural extension to computation of the
Holstein-$t$-$J$ model.

\acknowledgments One of the authors (J.B.) acknowledges  the warm
hospitality during his visit at the Institute for Materials
Research, Tohoku University, Sendai. J.B. also acknowledges
stimulating discussions with A. Ram\v sak, I. Sega, and P.
Prelov\v sek, M. Stout for providing editorial suggestions, J.
Vidmar for encouragement, and the financial support of the
Slovenian Research Agency under grant P1-0044. S.M. and T.T.
acknowledge the financial support of the Next Generation Super
Computing Project of Nanoscience Program, CREST, and Grant-in-Aid
for Scientific Research from MEXT.

\bibliography{manutj}

\end{document}